\documentclass[dvips]{amsart}
\usepackage{amsmath}%
\usepackage{amsfonts}%
\usepackage{amssymb}%
\usepackage{graphicx}%

\begin{document}
\title[QCD effective action]{Finiteness of the Coulomb gauge QCD perturbative effective action}
\author[Andrasi]{A. Andra\v si}
\curraddr[Andrasi]{Vla\v ska 58, Zagreb, Croatia}

\email[Andrasi]{aandrasi@irb.hr}

\author[Taylor]{J. C. Taylor}
\curraddr[Taylor]{DAMTP, University of Cambridge, Cambridge, UK}
\email[Taylor]{jct@damtp.cam.ac.uk}

\date{February 9, 2015}

\def\p{\textbf{P}}
\def\q{\textbf{Q}}
\def\r{\textbf{R}}
\def\a{\textbf{A}}
\def\k{\textbf{K}}
\def\u{\textbf{U}}
\def\w{\textbf{V}}

\begin{abstract}
At 2-loop order in the Coulomb gauge, individual Feynman graphs contributing to the effective
action have energy divergences. It is proved that these cancel in suitable combinations of
graphs. This has previously been shown only for transverse external fields. The
calculation results in a generalization of the Christ-Lee term which was
inserted into the Hamiltonian. 
\end{abstract}
\maketitle
\clearpage

\section{Introduction}
In QCD, the Coulomb gauge has some attractions. It is the only manifestly unitary gauge (no ghosts).
It may be convenient in practice  when there is a natural rest frame, as in thermal QCD or for heavy quarks.
It has been used in studies of confinement (see for example \cite{confinement}). It has been used in lattice
calculation, see for example \cite{lattice}), and in studies of infra-red behaviour (see \cite{lebellac}).
But, in perturbation theory, it has complications due to \textit{energy divergences}, that is Feynman integrals which are divergent over the internal energy integrals (for fixed spatial momenta). These divergences
appear at 1-loop order, but in this case are easily curable (see \cite{lineardivs}), most simply by using, instead of the Lagrangian formalism,  the phase-space, Hamiltonian
formalism, as we do in this paper.

At 2-loop order, there are more subtle energy divergences. These occur in divergent integrals of the form
\begin{equation}
\int dp_0 dq_0 \frac{p_0}{p_0^2-P^2+i\epsilon}\frac{q_0}{q_0^2-Q^2+i\epsilon}F(\p,\q,\r)
\end{equation}
(capital letters, $\p,\q,\r$ etc denote spatial momenta and $P\equiv |\p |$ etc). This problem is resolved if it can be shown
that graphs can be grouped so that integrals like (1) appear in the combination 
\begin{equation}
(1/3)\int dp_0 dq_0 dr_0\delta(p_0+q_0+r_0)F(\p,\q,\r)$$

$$\times \left[{\frac{p_0}{p_0^2-P^2+i\epsilon}}{ \frac{q_0}{q_0^2-Q^2+i\epsilon}}+      {\frac{q_0}{q_0^2-Q^2+i\epsilon}} \frac{r_0}{r_0^2-R^2+i\epsilon} +  {\frac{r_0}{r_0^2-R^2+i\epsilon} \frac{p_0}{p_0^2-P^2+i\epsilon}}         \right]$$
$$=-(\pi^2/3) F(\p,\q,\r)
\end{equation}
as may be shown by contour integration.

For definiteness, suppose that we are calculating, to 2-loop order, the effective action $\Gamma(A^a_i)$
as a functional of the external vector field $A_i^a$ ($i,j,..$ are spatial indices and $a,b,c,..$ denote colour). (Parts of $\Gamma$ depending on $A_0^a$ or $E_i^a$
contain no divergent integrals like (1).)

The requisite theorem, that graphs can be combined to give (2), was proved by Doust \cite{doust} (see also \cite{chengtsai}) for the special case
of transverse external fields, $\partial_iA_i^a=0$. But the effective acion is formally defined (see the Appendix)
for general $A_i^a$, and this is indeed  required in order to state the BRST identities for $\Gamma$.
The purpose of the present paper is to extend Doust's theorem to general $A_i^a$. The $O(g^4)$ case was already studied in \cite{aajct}.
\pagebreak
The divergent integrals (1) are connected with the fact that operator ordering of the
Hamiltonian (which is non-polynomial) may not be well-defined. Christ and Lee \cite{christlee} (and references therein)
argued that the correct ordering entailed adding onto the naive Hamiltonian
two operator functionals, called $V_1$ and $V_2$, of the operator $\hat{A}_i^a$, which in the Coulomb gauge
is transverse by definition. To avoid double counting, the Feynman integrals (1) and (2) would have to
be omitted. Doust found that the integrals (2) contributed the same functionals $V_1+V_2$ to the effective action.
\subsection{The flow gauge}
In order to manipulate integrals like (1) rigorously, a temporary regularization is necessary.
Dimensional regularization (which we use, with space-time dimension $d$, to control ordinary ultra-violet divergences) does not
have any effect on (1). So we (following Doust) employ a 'flow gauge' with a parameter $\theta$.
For finite $\theta$ the function $F$ in (1) depends also on $\theta p_0, \theta q_0, \theta r_0$ so as to make the integral converge. So in (1) and (2) we must replace $F$ by (we write $p$ for the Lorentz vector $(p_0;\p)$)
\begin{equation}
F(\p,\q,\r,\theta p_0, \theta q_0, \theta r_0)=F(p,q,r).
\end{equation}
The Coulomb gauge is the limit $\theta \rightarrow 0$, which we take only after getting the
combinations (2). ($\theta=1$ is the Feynman gauge, so the flow gauge is sometimes called
an interpolating gauge).
\pagebreak[1]
The flow gauge is defined by the gauge-fixing term
\begin{equation}
-\frac{1}{2\theta^2}[(\partial_i \hat{A}_i^2)^2-\theta^2(\partial_0\hat{A}_0^a)^2]
\end{equation}
(the limit $\theta \rightarrow 0$ imposes the transversality condition on the operator $\hat{A}_i^a$).
The effect of this is that the bare Coulomb propagator, $-1/P^2$, is everywhere replaced by
\begin{equation}
-\frac{1}{P^2-\theta^2p_0^2}\equiv -\frac{1}{\overline{P^2}}
\end{equation}proof.tex
This is  what causes the function $F$ in (1) and (2) to be replaced by (3).

In the Coulomb gauge, ghost loops exactly cancel closed Coulomb loops. But in the flow gauge
this is not exactly the case. Ghosts, but not Coulomb loops, have a coupling $O(\theta^2)$ to the Coulomb field.
Hence, as well as graphs similar to Fig.1, graphs with ghost loops have to be included in the proof.
When the convergent form (2) is finally attained, the ghost graphs can be omitted.

The method of our proof, following \cite{doust}, is to show that $F(p,q,r)$ is a cyclically symmetric function of its arguments. Then the integrand in (1) can be symmetrized, to give (2). Finally, having attained
the convergent combination in the square bracket in (2), the limit $\theta \rightarrow 0$ can be taken in F in (2) giving the Coulomb gauge.

\begin{figure}[h]
\centering
\includegraphics[width=0.8\textwidth]{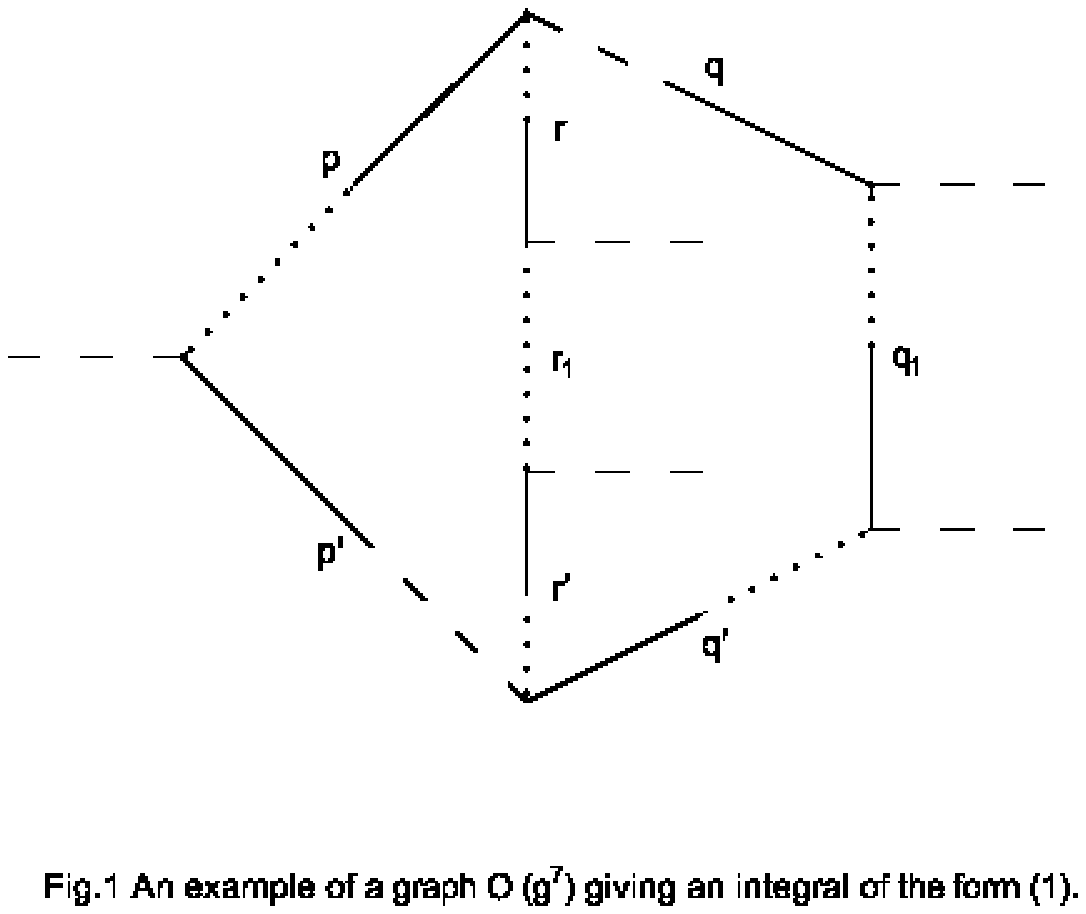}
\end{figure}
\begin{figure}[h]
\centering
\includegraphics[width=0.8\textwidth]{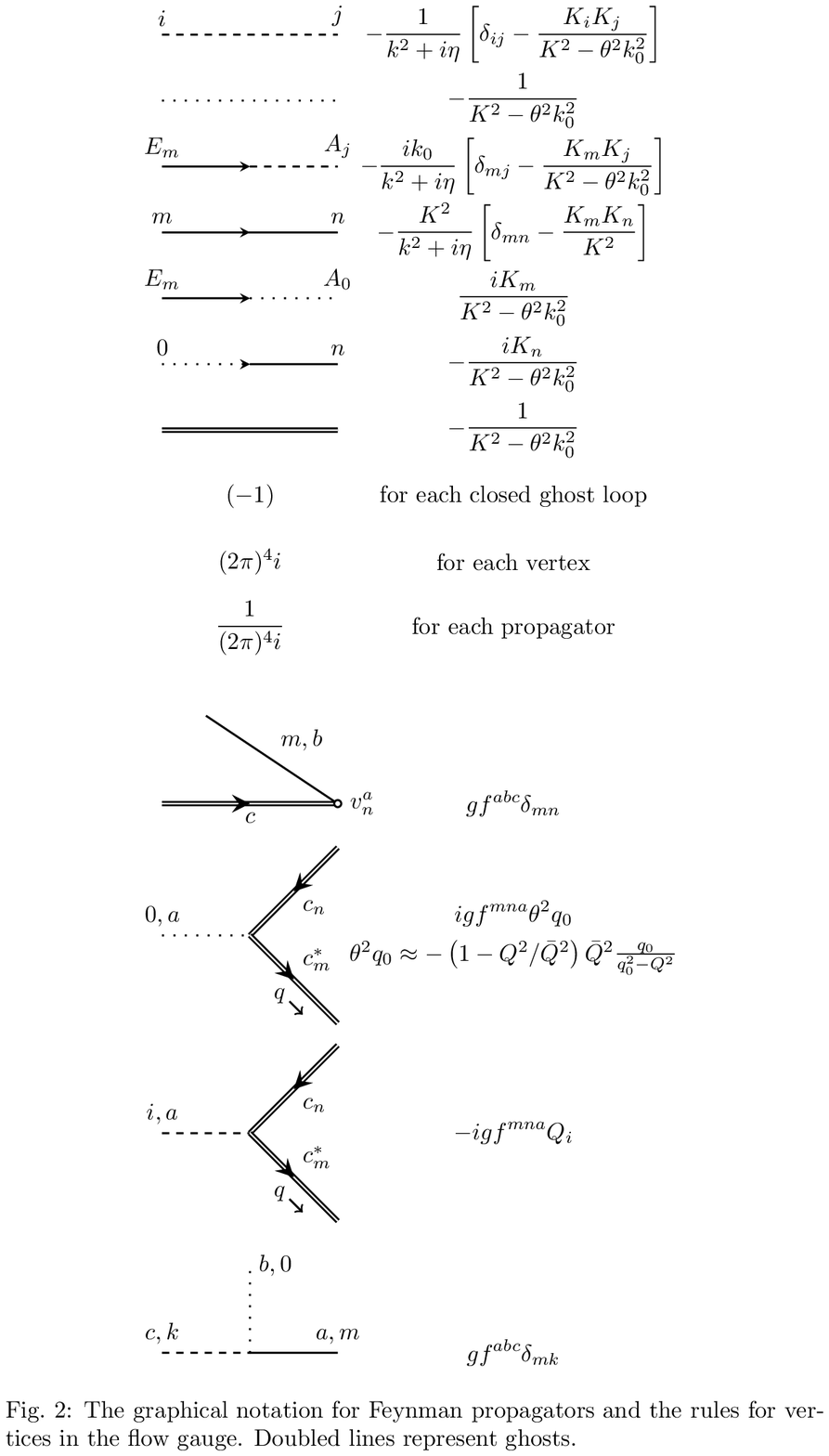}
\end{figure}

\subsection{An example}
An example (at $O(g^7)$) of a Feynman graph leading to an integral of the form of (1) is shown in Fig.1.
Our  notation for Feynman graphs is explained in  Fig.2. Dotted lines denote the propagator (5).
Dashed lines denote the propagator for $\hat{A}_i^a$ (transverse in the Coulomb gauge, but not transverse for
nonzero $\theta$). Solid lines always denote a   momentum (either spatial or time-like), multiplied by $i$, in the numerator (the sense of the momentum being directed towards the neighbouring vertex).
The external dashed lines represent the external field $A_i^a$. In Fig.1, the two denominators in (1)
come from the lines labeled $p'$ and $q$ in Fig.1.

In the Coulomb gauge ($\theta=0$) all the relevant graphs have a `spine' coming from the Coulomb
operator in the Hamiltonian. In Fig.1 this spine consists of the $p, r, r_1, r', q', q_1$ lines.
In addition to this spine, there are two lines ($q, p'$) where $\hat{A}_i$ and $\hat{E}_j$ operators in the Coulomb
operator have been contracted with one-another. In the flow gauge, the structure of the graphs are the
same though their interpretations are not as simple.

In Fig.1 there is just one line, labeled $r_1$, which has a dotted line with no solid line on it.
In all the relevant graphs there is on the spine just one line like this, which may be in any position,
on the centre line or the righthand or lefthand one. (To order $g^n$ there are $n-1$ such possible positions). In the case of transverse $A_i^a$, treated by Doust,
the contribution is independent of the position of this line, which simplifies the argument.
We show that the proof can be made without this simplifying property.

We need a compact notation in order to write down the general graphs giving integrals of the form of (1).
This we explain in the next section.
\begin{figure}[h]
\centering
\includegraphics[width=0.8\textwidth]{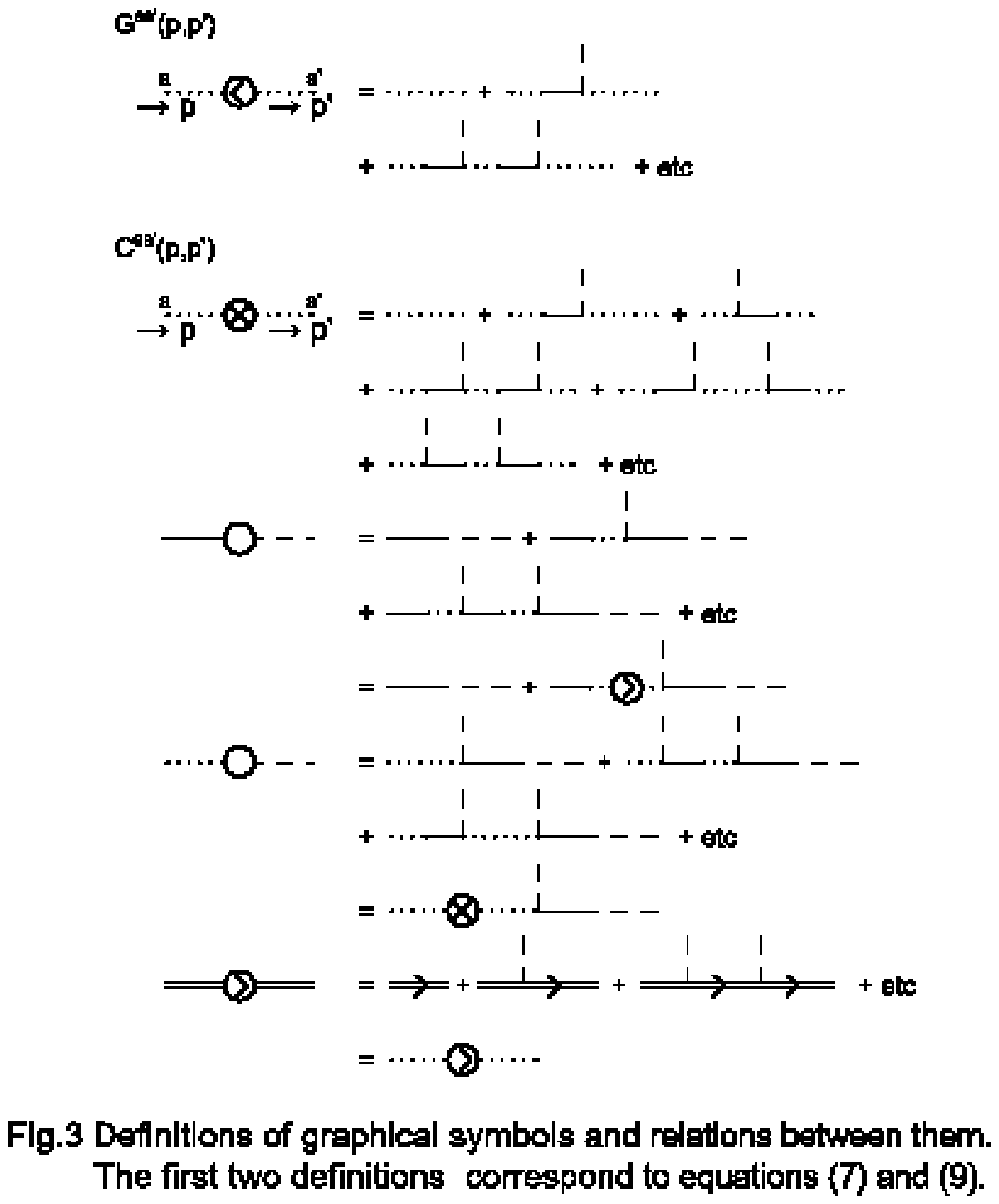}
\end{figure}
\section{The ghost and  Coulomb propagators}
The example in Fig.1 has three \emph{chains} of lines. Each chain consists partly of a propagator in the  external field
$A^a_i$. On the left and right in Fig.1, these propagators have the same form as the ghost propagator (though
they are not ghost lines), and in the centre the Coulomb propagator occurs. We need a compact notation
for these two propagators, which we now give.

\def\p{\textbf{P}}
\def\q{\textbf{Q}}
\def\r{\textbf{R}}
\def\a{\textbf{A}}
\def\k{\textbf{K}}
\def\u{\textbf{U}}
\def\w{\textbf{V}}
We call the ghost propagator $G^{aa'}(\p,\p';\theta p_0, \theta p'_0)$. In the Coulomb gauge, there would be no dependence
on $p_0,  p'_0$, but in the flow gauge, such dependence is introduced through the denominators (5). 

Define the  antisymmetric  matrix field
\begin{equation}
A_i^{ab}(k)=f^{abc}A_i^c(k)
\end{equation}
where $A_i^c(k)=A_i^c(-k)$ is the  external field in momentum space.

The ghost propagator $G$ is defined by the integral equation
\begin{equation}
-\overline{P^2}G^{aa'}(p,p')=\delta^{aa'}\delta^d(p-p')+ig\int d^dp_1 \p.\a^{ab}(p_1-p)G^{ba'}(p_1,p').
\end{equation}
This may be solved iteratively to give a series in $g$. As an example, the  second order term is
\begin{equation}
g^2\int d^d{p}_1[\overline{P^2}\overline{P_1^2}\overline{P'^2}]^{-1}\p.\a^{ab}(p_1-p)\p_1.\a^{ba'}(p_1-p').
\end{equation}

The Coulomb propagator $C^{aa'}(p,p')$ is given in terms of the ghost propagator by

\begin{equation}
C^{aa'}(p,p')=\int d^dp_1 G^{ab}(p,p_1)(-\p_1^2)G^{a'b}(-p',-p_1).
\end{equation}

As an example, one of the three second order terms in $C$ is
\begin{equation}
g^2\int d^d\p_1[\overline{P^2}\overline{P_1^2}\overline{P'^2}]^{-1}\p.\a^{ab}(p_1-p)\p'.\a^{a'b}(p'-p_1).
\end{equation}

The example (10) occurs on the centre chain in Fig.1 (with $p$ replaced by $r$).

Note the difference between (8) and (10) in general, but for transverse $A$ they are equal.

The series for $G$ and $C$ are illustrated graphically in Fig.3.
\begin{figure}[h]
\centering
\includegraphics[width=0.8\textwidth]{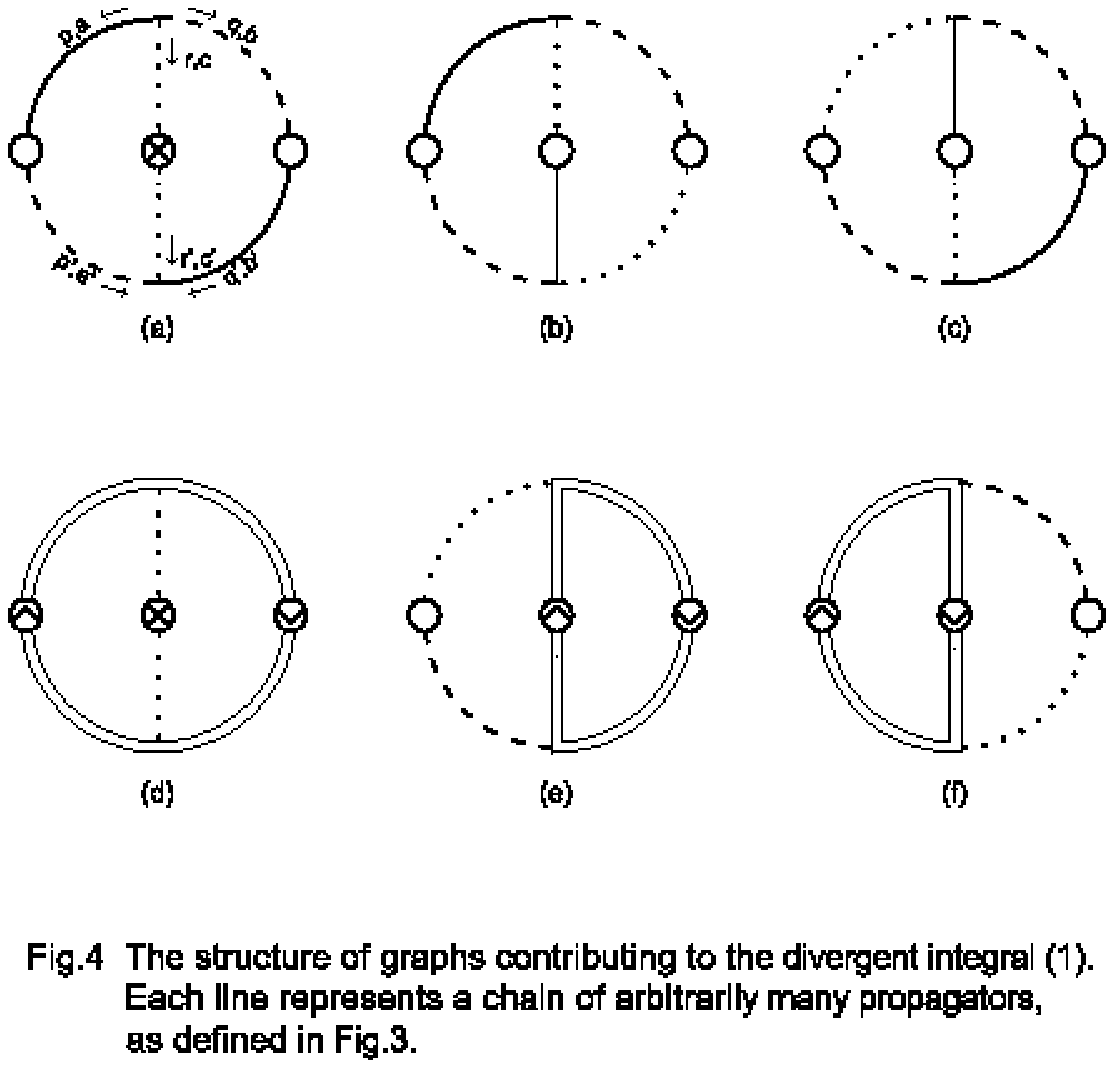}
\end{figure}
\section{The energy-divergent graphs}

We need a notation for graphs in which an arbitrary number of external gluon fields are
attached to the three chains. This notation is shown in Fig.3. Then the six types of graph which
give an integral containing (1) (with the substitution (5)) are shown in Fig.4. They are all drawn with 
the $p$-chain on the left and the $q$-chain on the right. These graphs give contributions of the form
\begin{equation}
\frac{1}{2}(2\pi)^{-d}f^{abc}f^{a'b'c'}\int d^dpd^dqd^dr\delta(p+q+r)d^dp'd^dq'd^dr'\delta(p'+q'+r')\left[\frac{p_0}{p^2}\frac{q'_0}{q'^2}\right]B,
\end{equation}
where the factors $B$ will be expressed graphically in Figs.7 and 8 later in this section.
(The factor $1/2$ in (11) is needed because the set of graphs which we use has symmetry under the interchange of the two internal vertices.)

In order to connect (11) to (1), we need a lemma.
This states that, up to terms $O(\theta$), we may replace $(p_0q'_0)/(p^2q'^2)$ or $p'_0q_0/(p'^2q^2)$
by 
\begin{equation}
\frac{p_0}{p^2+i\epsilon}\times\frac{q_0}{q_0^2+i\epsilon}
\end{equation}
(reminding ourselves of the $i\epsilon$ factors, which we usually leave understood).
The proof is as follows. The difference is an integral of the form
\begin{equation}
\int dp_0dq_0dr_0\delta(p_0+q_0+r_0)\left[\frac{p_0q'_0}{p^2q'^2} -\frac{p_0q_0}{p^2q^2}\right] F(\p,\q,\r;\theta p_0, \theta q_0, \theta r_0; \theta k^{(m)}_0,\k^{(m)})
\end{equation}
where the $k^{(m)}$, $(m=1,2,3..)$ are external  gluon momenta. Also $q'_0-q_0$ is equal to a linear combination
of a subset of the $k^{(m)}_0$. We now make a change of variables
\begin{equation}
\breve{p}_0=\theta p_0, \breve{q}_0=\theta q_0, \breve{r}_0=\theta r_0.
\end{equation}
Then, in terms of the new variables of integration, the difference between the two terms in the square
bracket in (13) lies in dependence on $\theta^2 P^2, \theta^2 P'^2, \theta k_0^{(m)}$. The integral
is convergent, because of the terms in $F$ like 
\begin{equation}
\frac{1}{\breve{r}_0^2-\theta^2 \r^2};
\end{equation}
so we may let $\theta\rightarrow 0$, and the integrand in (13) is zero in the limit.

In view of this lemma, the square bracket in (11) may be replaced by (1).
Having done this, the  contributions to $B$ in (11) may, with the use of the definitions in Figs.3 and 5, be written as in Fig.7.
All these graphs have at least one external field attached to the $p$-chain and at least one to the $q$-chain
(these are the external dashed lines shown explicitly; in addition there are arbitrarily many attachments
in the $C$ and $D$ propagators). But there are the special cases in which there is no external attachment to the $p$-chain, or none to the $q$-chain, or none to either. Some examples of these cases are shown in
Fig.8.
\begin{figure}[h]
\centering
\includegraphics[width=0.8\textwidth]{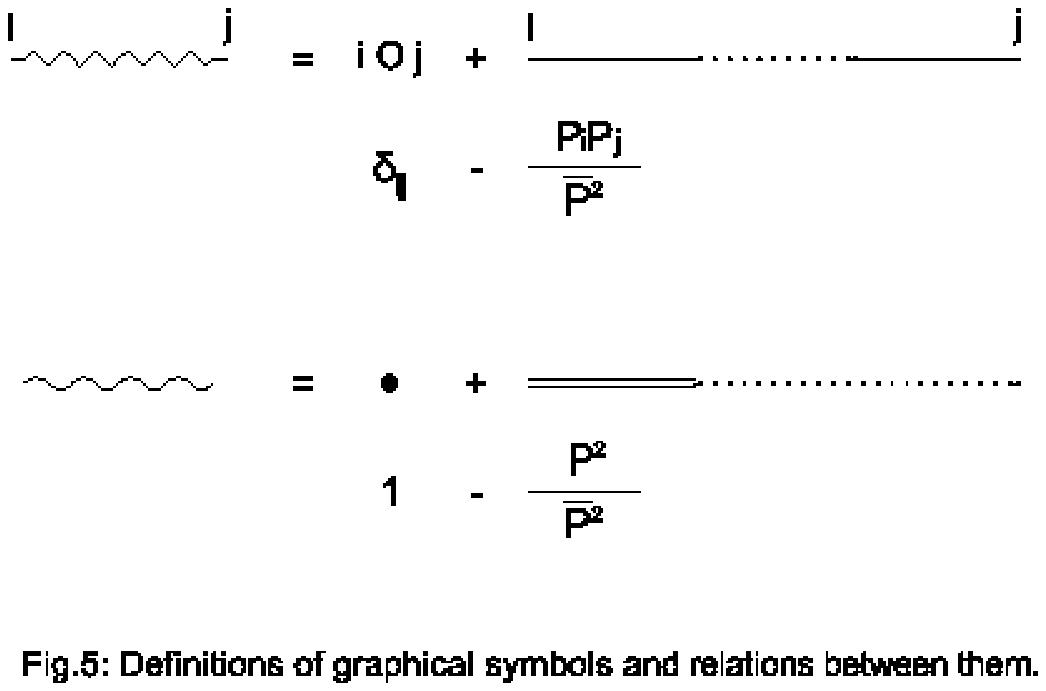}
\end{figure}
\begin{figure}[h]
\centering
\includegraphics[width=0.8\textwidth]{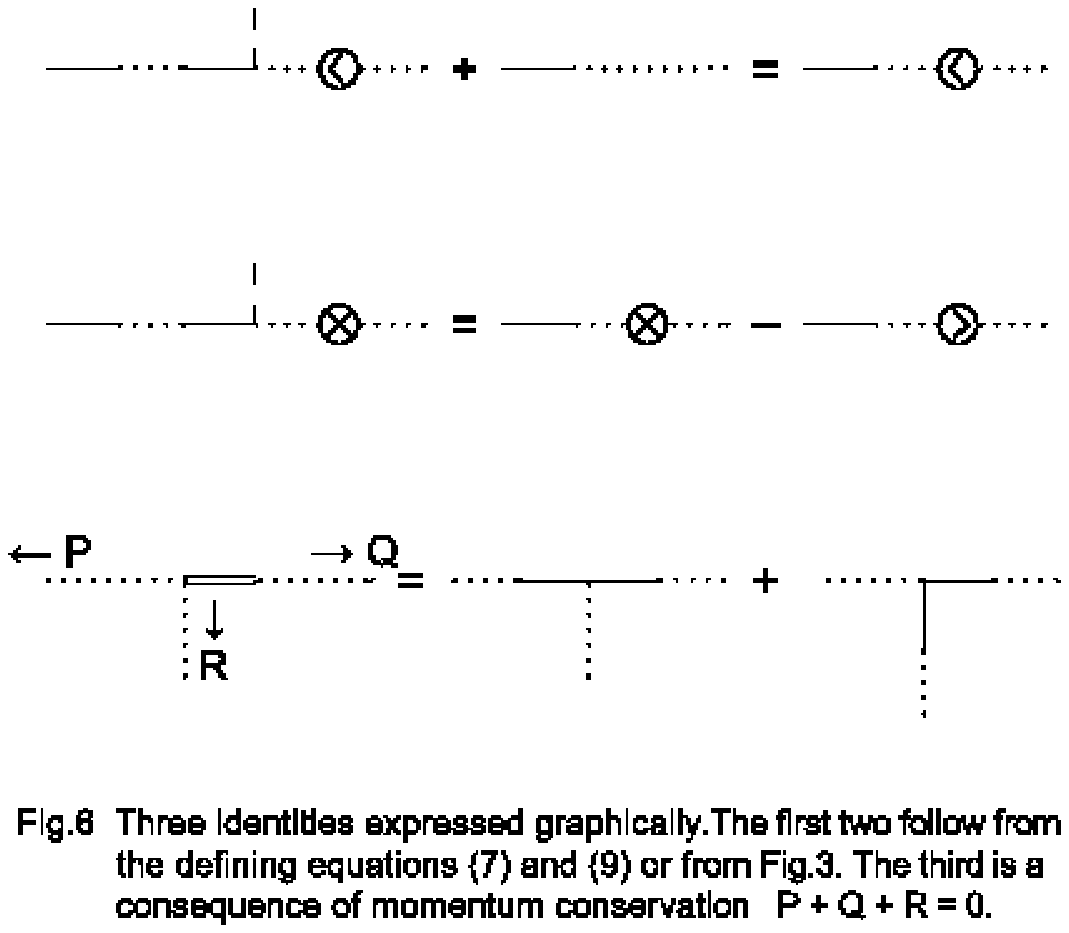}
\end{figure}
Our aim is to show that the sum of the graphs in Figs.7 and 8 has symmetry under cyclic permutations of
\begin{equation}
(p,p';a,a'),\\(q,q';b,b'),\\(r,r',c,c').
\end{equation}
 Then, with the integration (11), (1) can be replaced by the convergent
integral (2).

To order $g^n$, there are contributions with either $n+1$ or $n$ or $n-1$ denominators like (5).
It turns out that each of these three classes separately has cyclic symmetry; so we treat them
one by one in the following subsections.
\begin{figure}[h]
\centering
\includegraphics[width=0.8\textwidth]{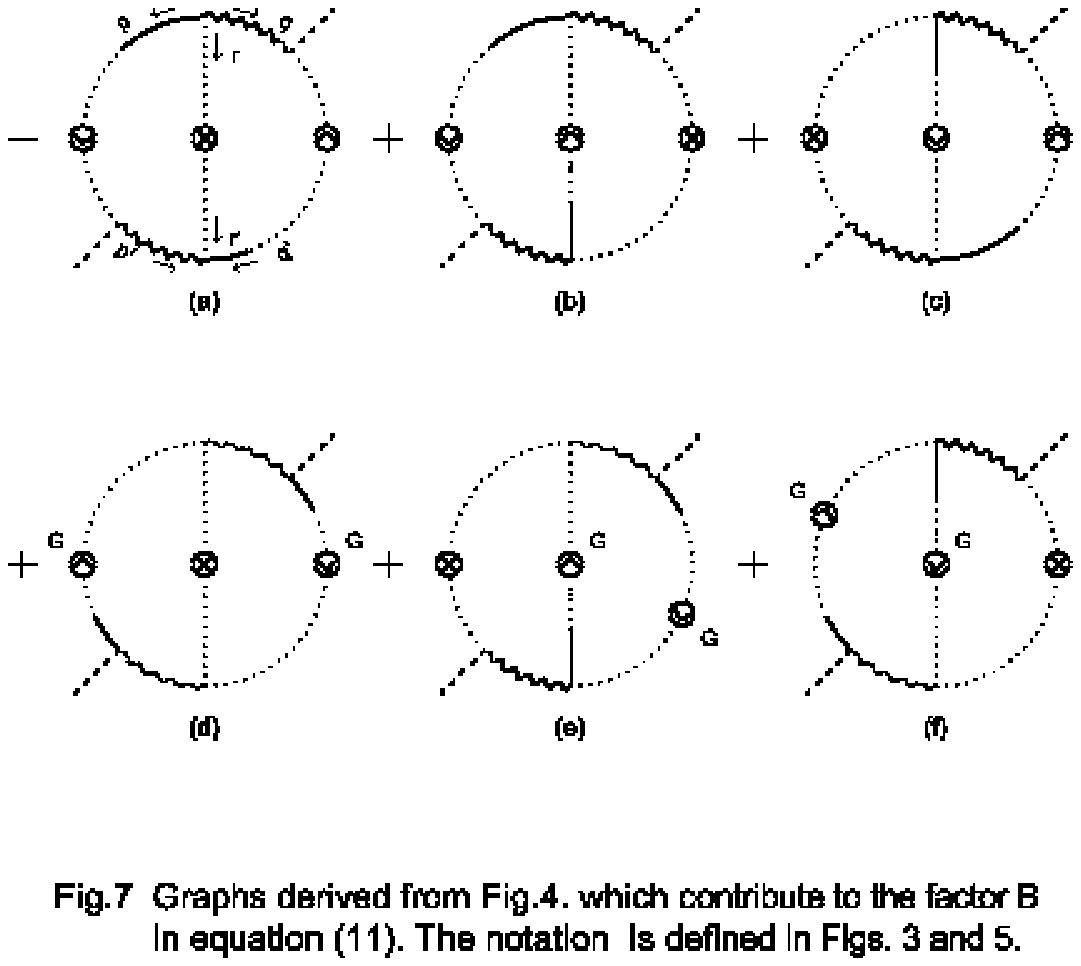}
\end{figure}
\subsection{\texorpdfstring{$(n+1)$}{n+1 denominators} denominators}
\begin{figure}[h]
\centering
\includegraphics[width=0.8\textwidth]{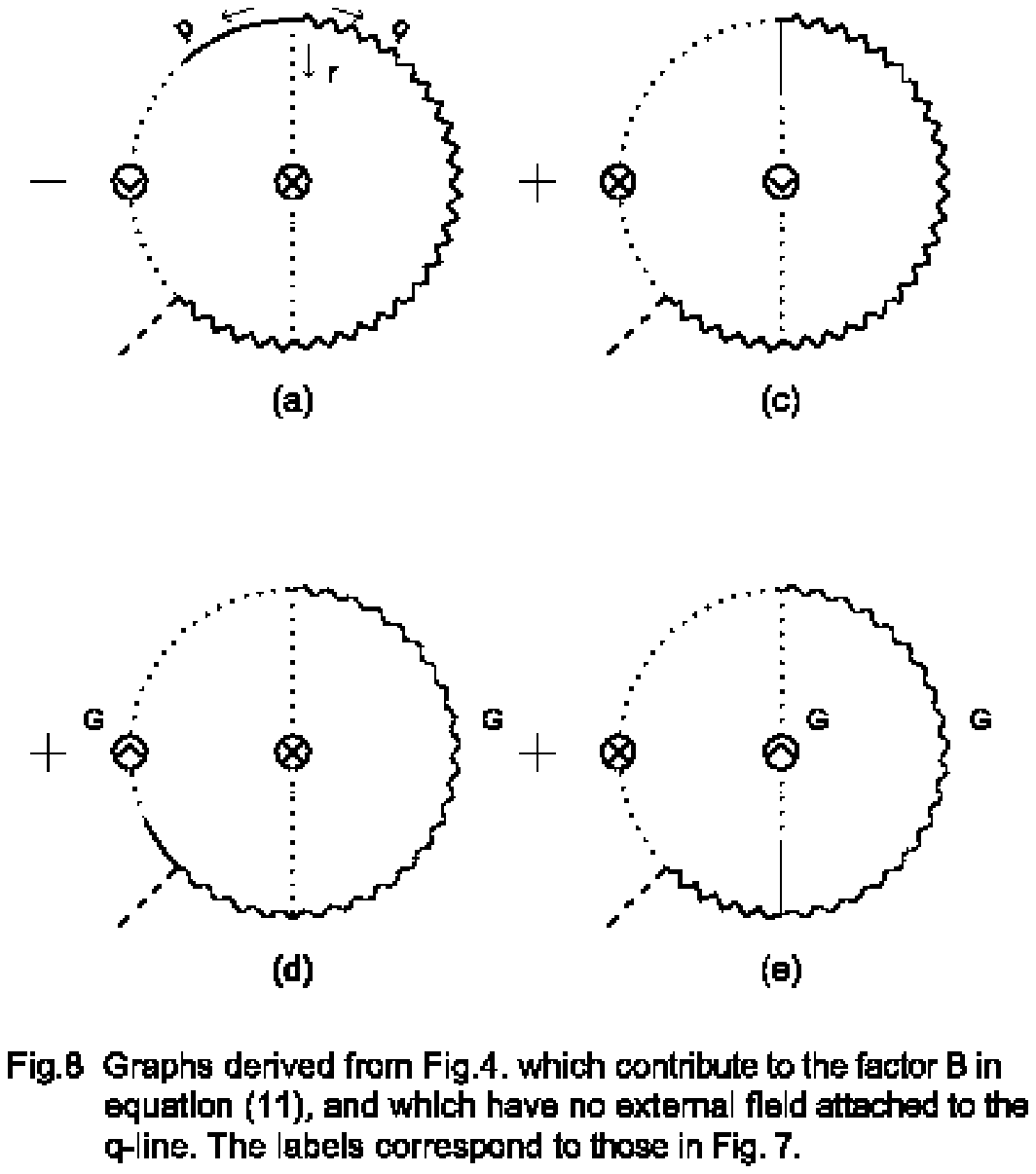}
\end{figure}
For these terms, in the graphs in Fig.7, we take the $P_iP_j/\bar{P^2}$  and $P^2/\bar{P^2}$ terms
from  Fig.5. We then use  the first two identities in Fig.6, which follow from (7) and (9),
thus getting one contribution from (a) and two each from (b), (c), (d), (e) and (f). In the ghost graphs (d), (e) and (f), we  use also the third identity in Fig.6, which follows from momentum conservation. As a result of these steps, there are finally four contributions from each of (d), (e) and (f). The total number of contributions from all six graphs is
thus 17. There are sign factors. The ghost graphs (d), (e), (f) each have a minus sign. There are additional minus signs for (a), (d), (e) and (f) because there colour matrices are in a different order from that in  equation (11). Finally, contributions from the last term in the second identity in Fig.6 receive another minus sign.
\begin{figure}[h]
\centering
\includegraphics[width=0.8\textwidth]{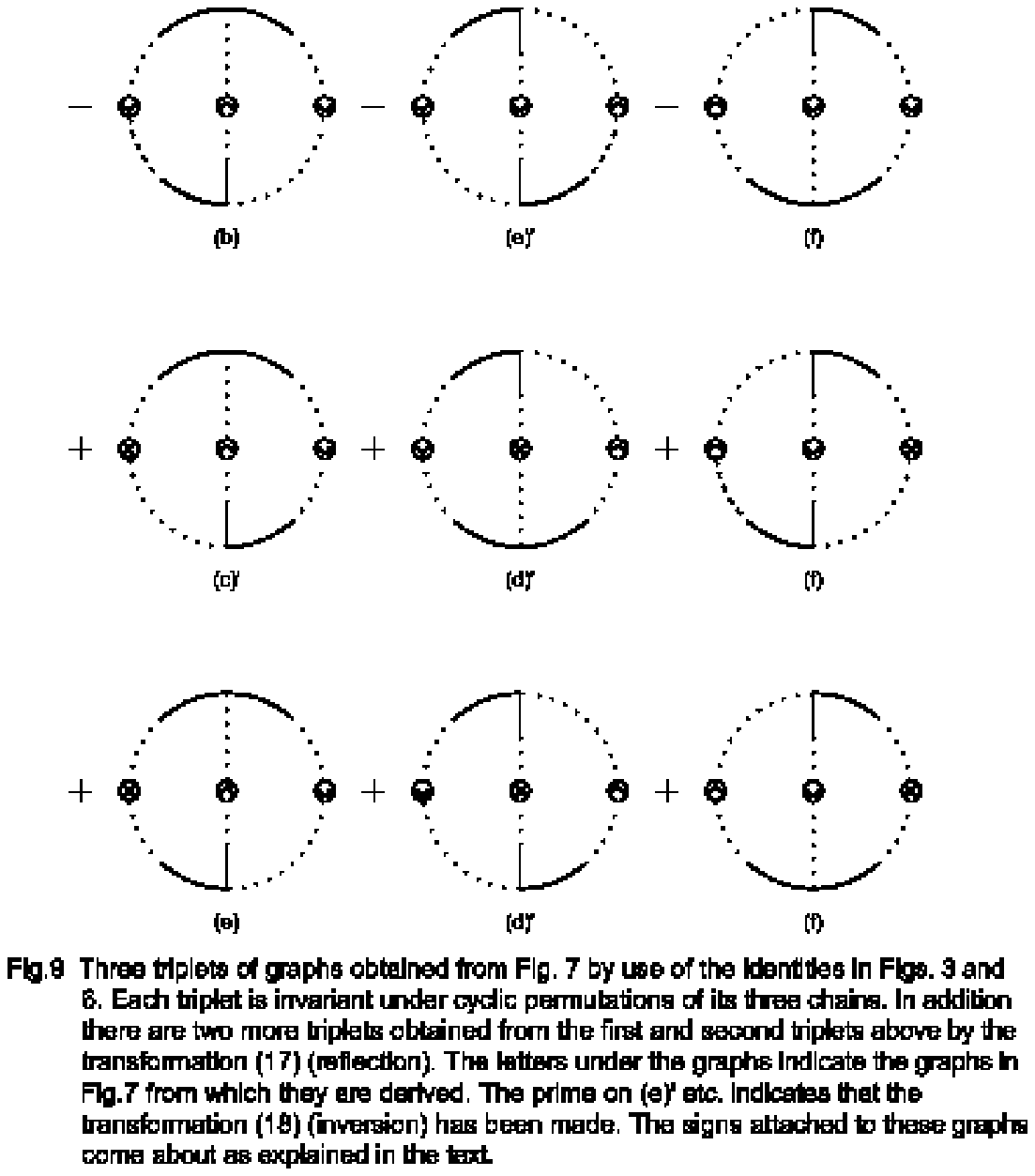}
\end{figure}
\begin{figure}[h]
\centering
\includegraphics[width=0.8\textwidth]{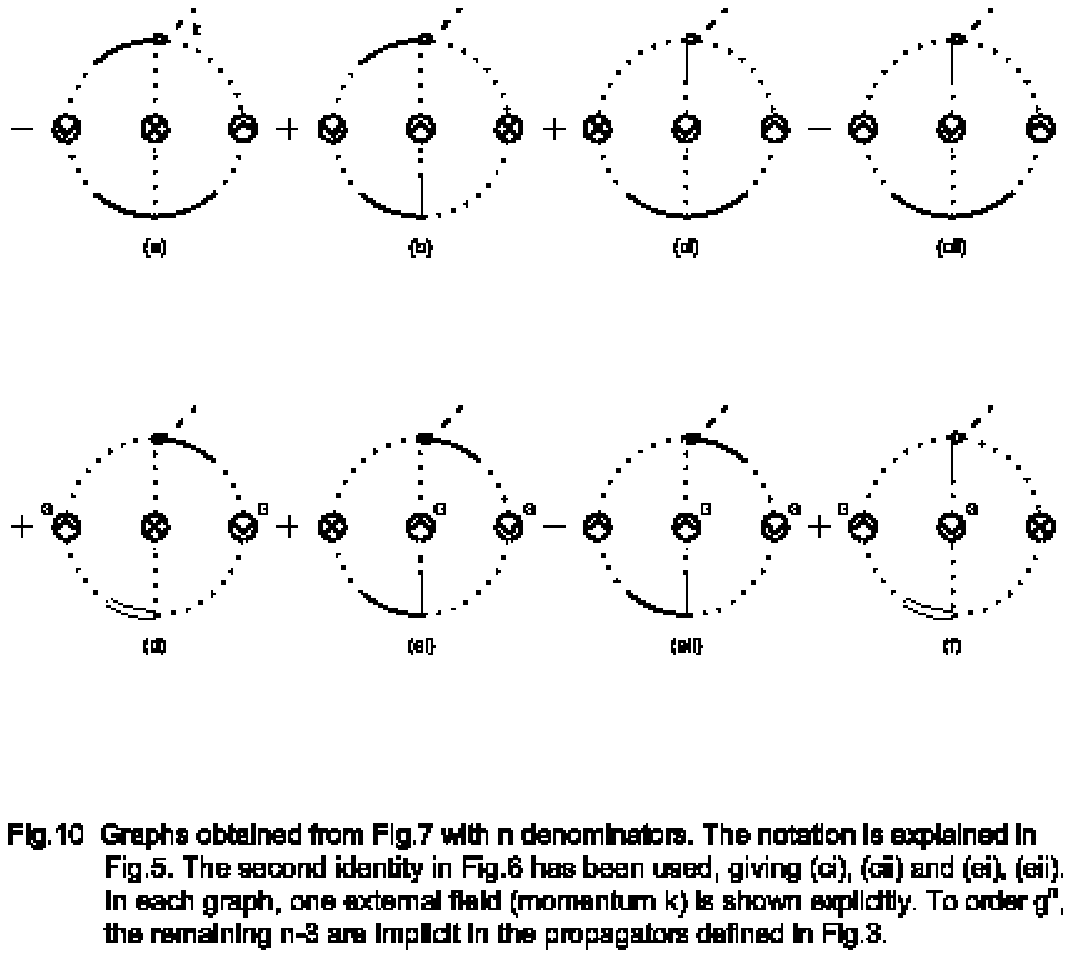}
\end{figure}
The term from (a) is canceled by one of the contributions from (d), so that fifteen graphs remain.
These fall into five triplets, each triplet being invariant under cyclic permutations of the three chains, that is cyclic permutations of the variables in (16). This allows the divergent integral (1) to be replaced by the convergent
integral (2). Three of the triplets are shown in Fig.9. The remaining two are obtained simply by
reflection, that is the change of variables
\begin{equation}
p,p',q,q',r,r';a,a',b,b',c,c'\leftrightarrow q,q',p,p',r'r';b,b',a,a',c,c'.
\end{equation}

In some of the graphs in Fig.9 we have made use of the change of variables
\begin{equation}
p,q,r; a,b,c\leftrightarrow -p', -q', -r';a',b',c'
\end{equation}
that is inversion of the graph. When this has been done it is indicated by a prime, e.g. $(c)\rightarrow (c)'$

In special cases, some graphs from Fig.8 have to be added to those from Fig.7, in order to contribute the last
term in the first identity in Fig.6.
\begin{figure}[h]
\centering
\includegraphics[width=0.8\textwidth]{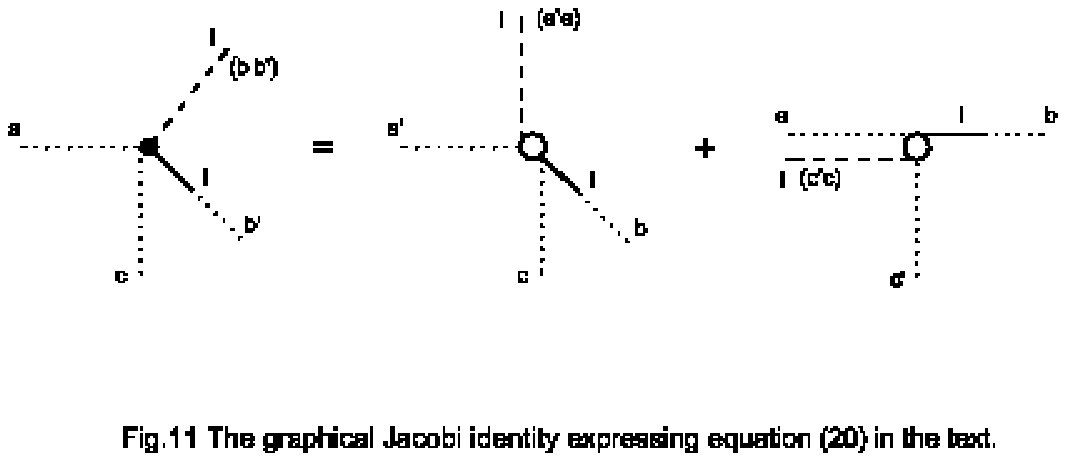}
\end{figure}
\begin{figure}[h]
\centering
\includegraphics[width=0.8\textwidth]{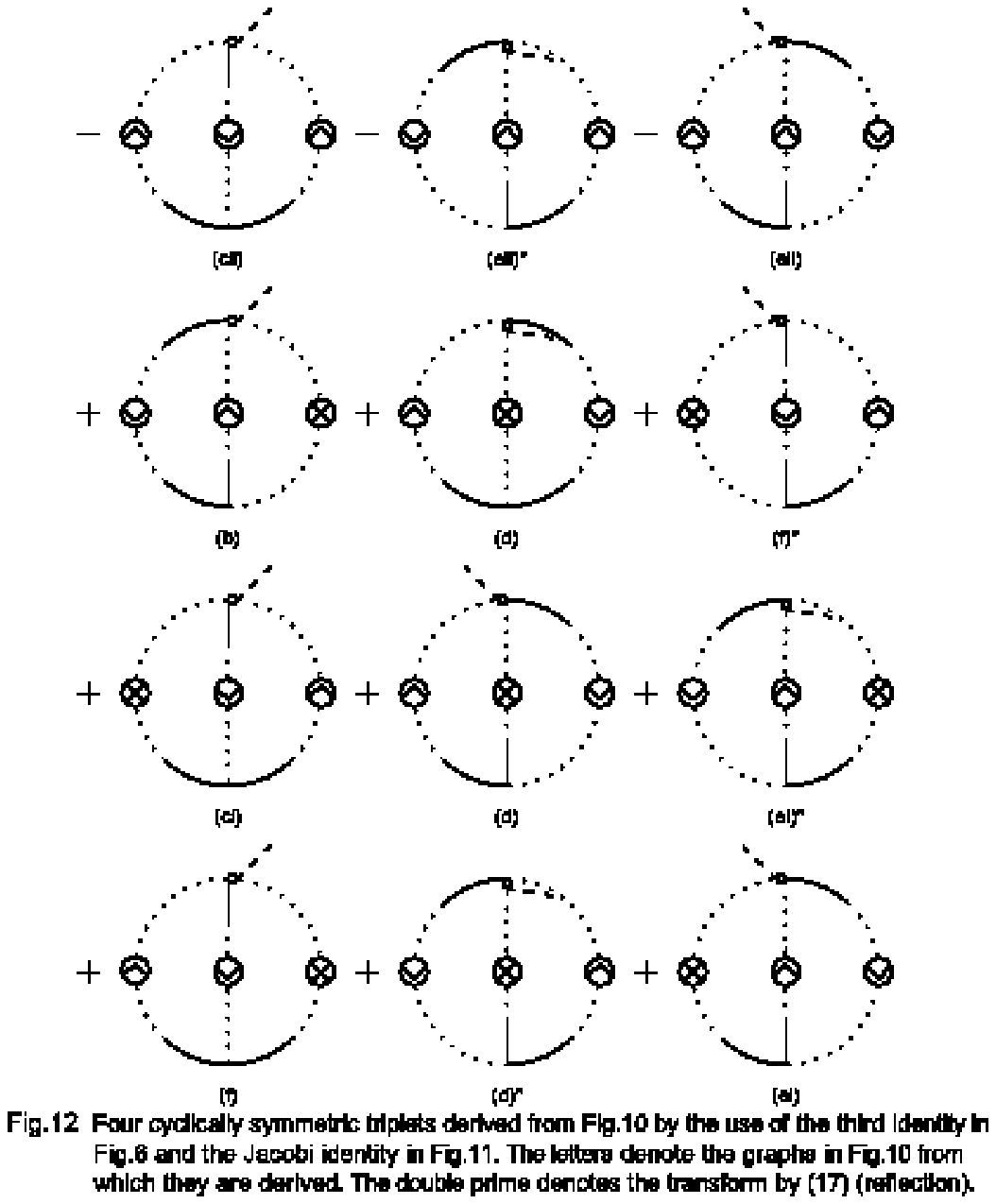}
\end{figure}
\subsection{\texorpdfstring{$n$}{n denominators} denominators}

For these terms, we take from Fig.7 just one term with $\delta_{ij}$ or with $1$. Because of symmetry under (18) (inversion), we can wihtout loss of generality take these to be at the top of the graphs. Then the possible graphs are shown in Fig.10, with the notation in Figs.3 and 5. 
At the top vertex, momentum conservation takes the form
\begin{equation}
\int d^dk  \delta^d(p+q+r-k)A_i^{bb'}(k),
\end {equation}
where $A_I^{cc'}(k)$ is the external field shown explicitly by the dashed lines in Fig.10. Expression (19) is symmetric in $p,q,r$.

 In Fig.10, use has already been made of the first identity in Fig.6. To simplify these graphs, we use the third identity in Fig.6 for graphs (d) and (f), and apply Jacobi identity, shown graphically in Fig.11, to graphs (d), (ei) (eii). This identity, stated algebraically, is
\begin{equation}
f^{abc'}A_i^{cc'}=f^{ab'c}A_i^{b'b}+f^{a'bc}A_i^{a'a},
\end{equation}
where $A_i^{cc'}$ is defined in (6).
There result 14 graphs. Graph (a) cancels one of the contributions from (d), when we make use
of (17).  Again using (17) where necessary, we obtain 12 graphs, which fall into four groups of three, each group having cyclic symmetry. 
These are shown in Fig.12, labeled with the names of the graphs in Fig.10 from which they originate. Again, the cyclic symmetry enables us to replace  the divergent (1) by the convergent (2).
\begin{figure}[h]
\centering
\includegraphics[width=0.8\textwidth]{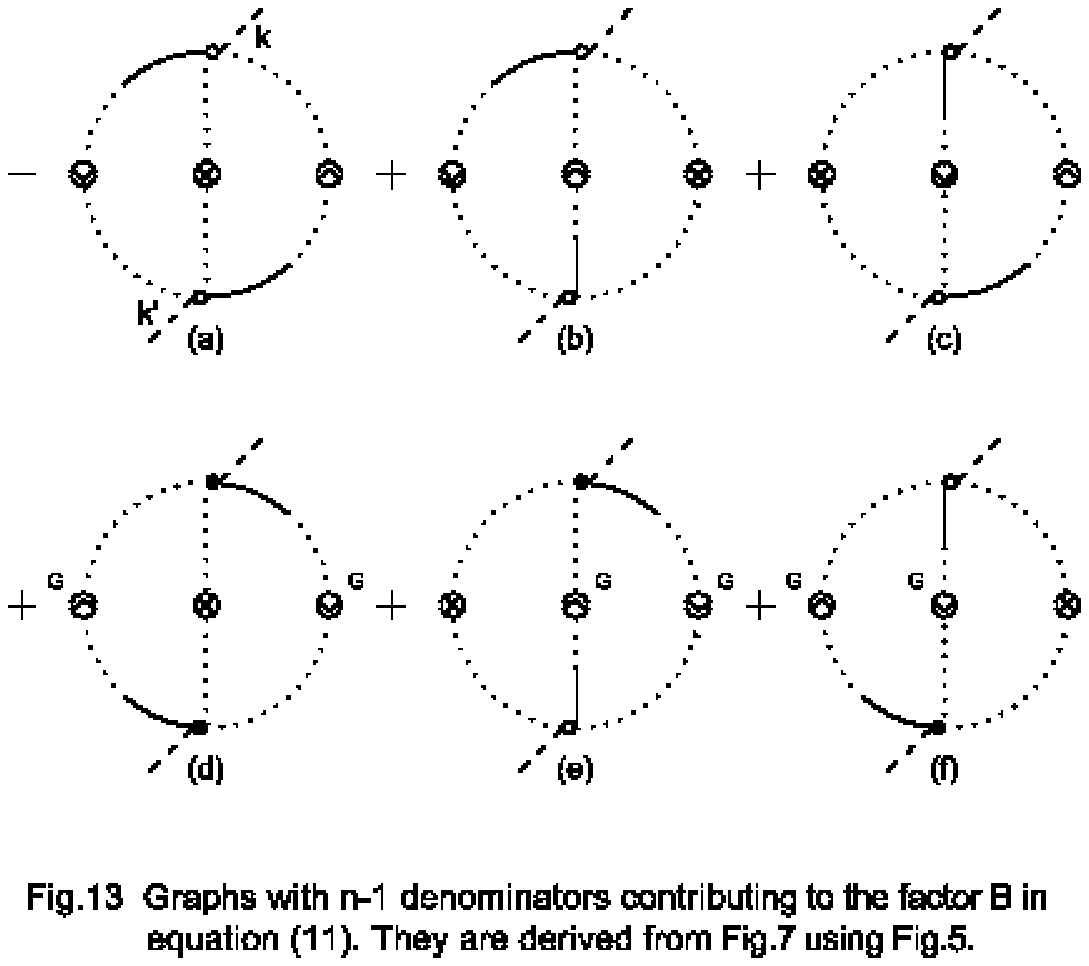}
\end{figure}
\begin{figure}[h]
\centering
\includegraphics[width=0.8\textwidth]{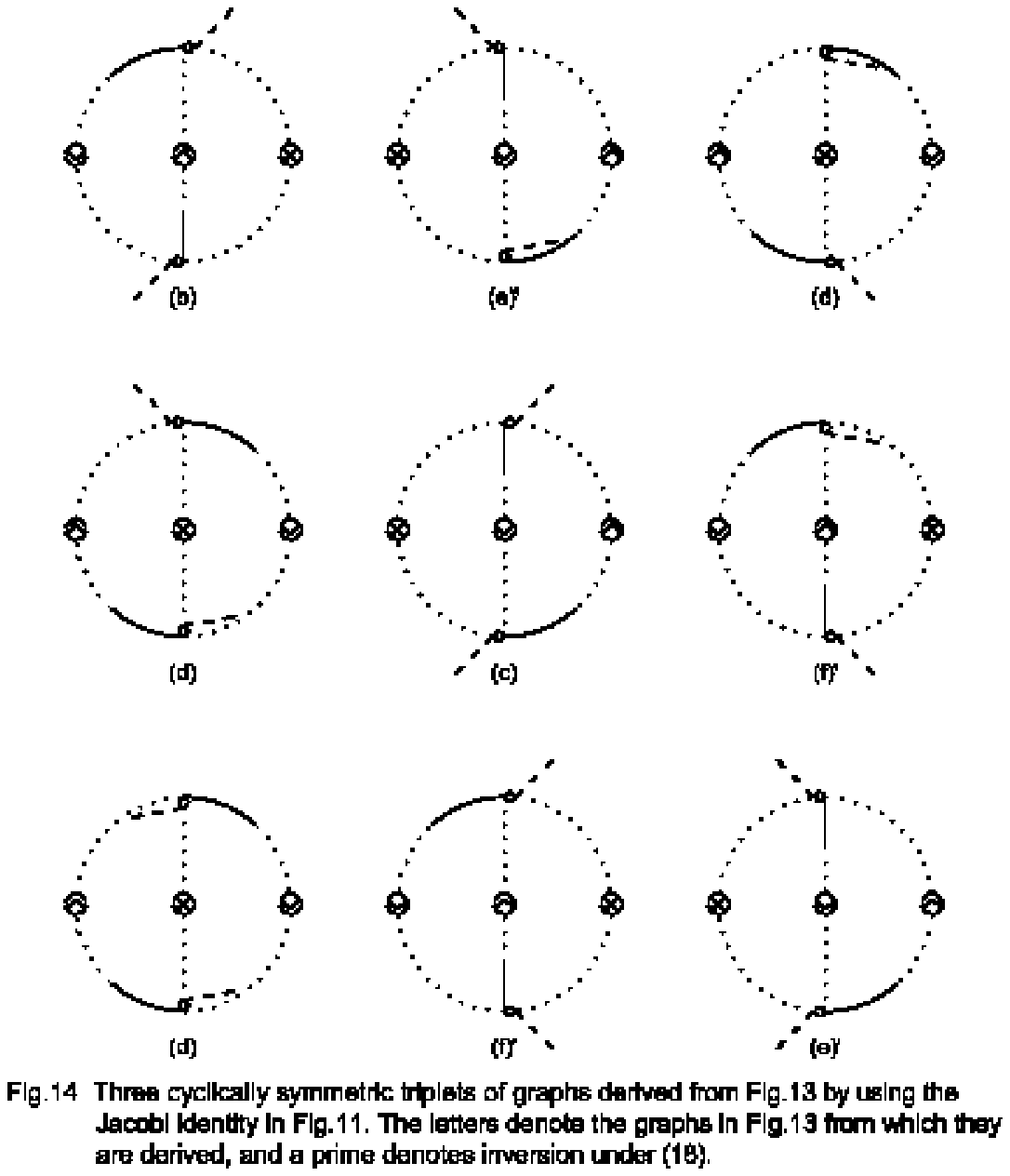}
\end{figure}
\subsection{\texorpdfstring{$(n-1)$}{n-1 denominators} denominators}

In this case, we take two terms $\delta_{ij}$ or $1$ from Fig.5 in Fig.7.
The results are shown in Fig.13. Then we use the Jacobi identity Fig.11 twice in graph (d) and once each
in graphs (e) and (f). There result 11 graphs. Graph (a) cancels one of the contributions from (d),
leaving nine graphs. These fall into three cyclically symmetric triplets, as shown in Fig.14, having used
the inversion symmetry in (18) where necessary. Once again, the cyclic symmetry enables the divergent
(1) to be replaced by the convergent (2) in (11).

This completes the proof for graphs in which at least one external field $A$ is attached
to each chain. But there are exceptional graphs for which this is not the case, like those in Fig.8.
We must now deal with these exceptional cases.
\begin{figure}[h]
\centering
\includegraphics[width=0.8\textwidth]{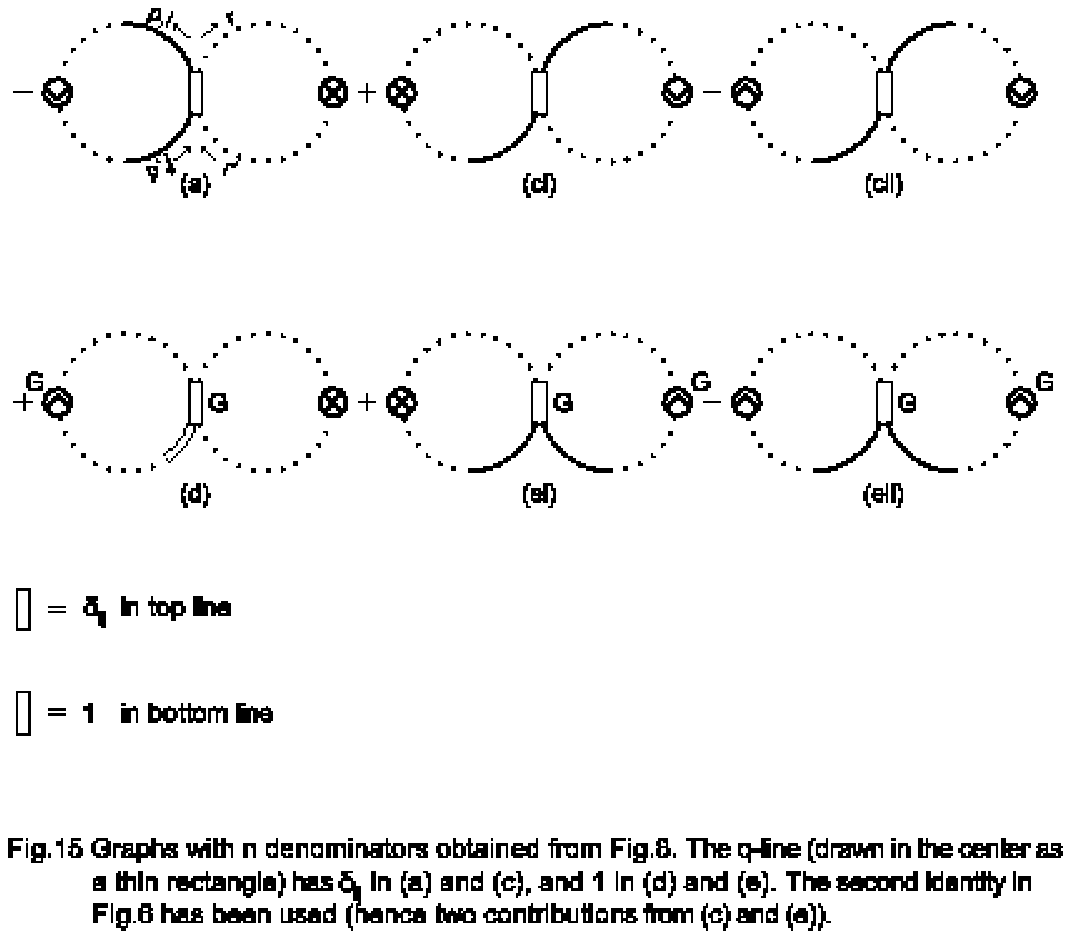}
\end{figure}
\section{Exceptional graphs}
These are graphs with either no external gluon attached to the $q$-chain or none attached to the $p$-chain.
Because of reflection symmetry (17), we can without loss of generality take the former case.
The graphs are shown in Fig.8. Again we have to consider the cases with $(n+1)$, $n$, or $(n-1)$
denominators, but the first of these is already included in the previous section, so we need only
consider the latter two cases.
\begin{figure}[h]
\centering
\includegraphics[width=0.8\textwidth]{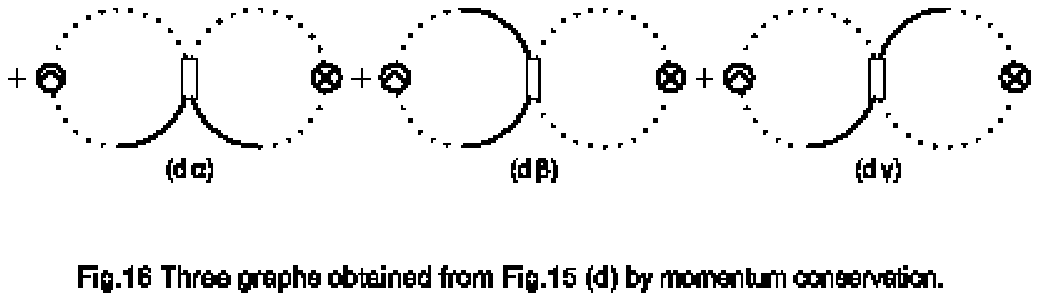}
\end{figure}
\subsection{\texorpdfstring{$n$}{n denominators} denominators}
The relevant graphs are shown in Fig.15, where the central rectangle denotes $\delta^{bb'}$ in the graphs (d), (e), (f); and $\delta^{bb'}\delta_{ij}$ in (a), (b), (c). The graphs $(ci), (cii), (ei), (eii)$   arise because of the use of the
second identity in Fig.6. The last identity in Fig.6 (momentum conservation)  is applied to graph (d) in Fig.15  to give the three graphs in Fig.16.

We then prove symmetry under the interchange of the $p$- and $r$-chains. Graph (a) in Fig.15 cancels graph (dii)   in Fig.16. Graphs $(cii)$ and ($eii$) in Fig.16 each separately have the required symmetry, making use
of inversion invariance under (18). And the two pairs
\begin{equation}
\{(ci)',(d\gamma)\};\\\\\\\\\ \{(ei),(d\alpha)\}
\end{equation}
each are invariant under the interchange.

This means that in (1), 
\begin{equation}
F(p,q,r)=F(r,q,p)
\end{equation}
($F(\p,\q,\r)$ becomes $F(p,q,r)$ in the flow gauge, as in (3)).
Therefore we can replace (1) by
\begin{equation}
(1/2)\int dp_0 dq_0 dr_0\delta(p_0+q_0+r_0)F(p,q,r)$$

$$\left[\frac{p_0}{p_0^2-P^2+i\epsilon} \frac{q_0}{q_0^2-Q^2+i\epsilon}+      \frac{q_0}{q_0^2-Q^2+i\epsilon}\frac{r_0}{r_0^2-R^2+i\epsilon}           \right].
\end{equation}

We note also that, in Figs.(15) and (16), $F$ factorizes
\begin{equation}
F= \phi_i^{aa'}(\p,\theta^2 p_0^2)\phi_i^{cc'}(\r,\theta^2 r_0^2),
\end{equation} 
neglecting terms $O(\theta k_0^{(m)})$ where $k^{(m)}$ are external momenta;
so
\begin{equation}
\int dp_0 \frac{p_0}{p^2}\phi_i^{aa'}=O(\theta),
\end{equation}
and similarly for the $r$-integral (just as at one loop order). This allows us to insert the third term $(p_0r_0/p^2r^2)$ into the square bracket in (23) and then use the identity in (2) to get
\begin{equation}
-(\pi^2/2)F
\end{equation}
and finally let $\theta \rightarrow 0$.
Thus we have achieved our aim, but there is the important difference from section 3 that we here have a factor
of a half rather than a third as in (2).
\begin{figure}[h]
\centering
\includegraphics[width=0.8\textwidth]{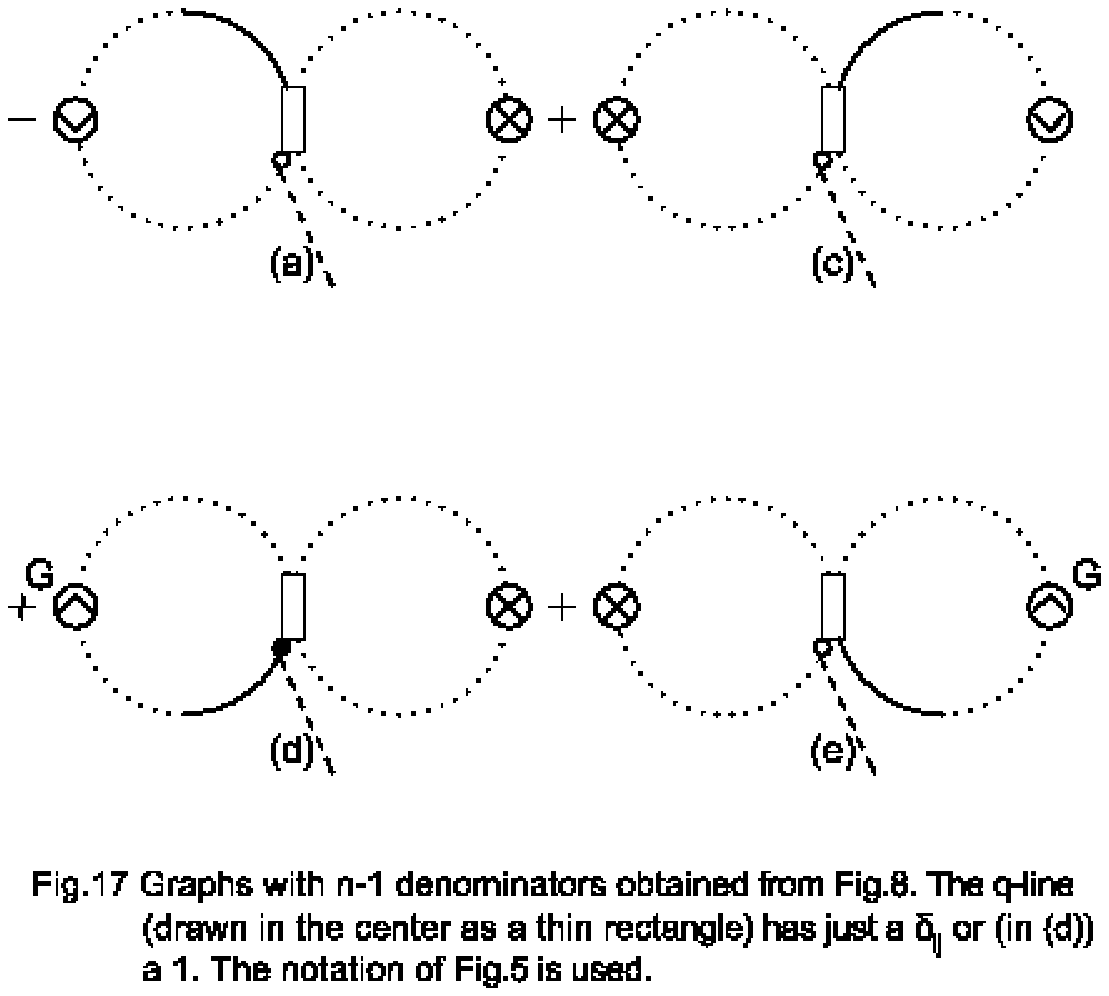}
\end{figure}
\subsection{\texorpdfstring{$(n-1)$}{n-1 denominators} denominators}

The relevant graphs are in Fig.17. For graph (d), we use the Jacobi identity (Fig.11) twice to get the three graphs in
Fig.18. As usual, graph (a) is canceled by (d1)'. The remaining four graphs fall into two
pairs
\begin{equation}
\{(e),(d3)\};\\\\\\\\\\\ \{(c)',(d2)\}
\end{equation}
each of which has symmetry under the interchange of the $p$- and $r$-lines. After that the argument
continues exactly as in subsection 4.1.

There is a special case in which, in Fig.4, both the $p$-chain and the $q$-chain in graph (a) consist of the
single terms $\delta_{ij}$, and in graph (d) of the single factors $1$. The result is just the same as in \cite{doust}, and is
  \begin{equation}
	\pi^2(d-2)\int d^{d-1}\p d^{d-1}\q d^{d-1}\r \delta^{d-1}(\p +\q+\r)f^{abc}f^{abc'}C^{cc'}(\r,\r').
	\end{equation}
	In dimensional regularization this integral is zero, but we write it here for completeness. Below we neglect all such terms which vanish in dimensional regularization.
	
	\section{The generalized Christ-Lee functions}
	
	In both section 3 and section 4, the energy integral has been shown to be finite, and could be done using equation (2). But in section 3 this resulted in a factor $-\pi^2/3$ while in section 4 the factor was
	$-\pi^2/2$ (leaving aside (28)). The limit $\theta \rightarrow 0$ could be taken. We present the results as the sum of two contributions. For the first we will  add together the contributions from sections 3 and 4
	with a factor $-\pi^2/3$ for both. In the second contribution, we compensate by using the result of
	section 4 but with a factor $-\pi^2/6$.
	
	\subsection{Contributions from Figs.7, 8}
	We take the former contribution first,  combining all the terms arising from the graphs in Figs.7 and 8. Since we have complete contributions (not separating Fig.8 from Fig.7) the ghost graphs disappear in the limit
	$\theta \rightarrow 0$, because of the factors $[1-(P^2/\bar{P^2})]$ in Fig.5. Thus the result comes from graphs (a), (b), (c) in Fig.7 and (a), (c) in Fig.8. It is straightforward to write these out using the rules in Figs.3 and 5. 
	
	We first note that, with $\theta=0$, the only external momentum dependence in (11) comes from the second $\delta$ function:, which we may write
	\begin{equation}
	\delta(p'_0+q'_0+r'_0)=\delta\left[\sum_m k_0^{(m)}\right]=\frac{1}{2\pi}\int dt \exp \left(-it\sum_m  k_0^{(m)}\right)
	\end{equation}
	where the $k^{(m)}$ ($m=1,2,3...$) are the momenta in the external fields. The effect of this is that
	the external field $A_i^a(\k,k_0)$ is everywhere replaced by the time Fourier transform
	$\tilde{A}^a_i(\k,t)$. From now on, that external fields in (7), (9) and (31), (32), (33) below are to be understood as meaning this.
	
	Then the contributions have the form
	\begin{equation}
	-(g^2/8)(2\pi)^{-d+1} f^{abc}f^{a'b'c'}\int dt\int d\p d\q d\r d\p' d\q' d\r' \delta(\p +\q +\r)\delta(\p' +\q'+ \r')F,
	\end{equation}
	where the terms in $F$   come from the graphs mentioned above. (In (30), the integrals and delta functions are understood to be $(d-1)$-dimensional.) Expression (30) is for the effective action; 
	for an effective Hamiltonian, the time integral and the minus sign should be omitted.
	
	Define
	\begin{equation}
	J^{aa'}_{ii'}(\p,\p')=T_{ii'}(\p)\delta^{aa'}\delta(\p-\p')$$
	$$+\int d\p'' T_{i''i}(\p)A_i^{aa''}(\p-\p'')G^{a''a'}(\p'',\p')(iP'_{i''})
		\end{equation}
		and
		\begin{equation}
		L^{aa'}_{i}(\p,\p')=\int d\p''T_{ii''}(\p) A_{i''}^{a''a}(\p-\p'')C^{a''a'}(\p'',\p')
		\end{equation}
	where the transverse projection operator $T_{i''i}(\p)=\delta_{i''i}-(P_iP_{i''})/P^2)$ comes from Fig.5 when
	$\theta =0$. With this notation Fig.7(a),(b),(c) together with Fig.8(a),(c)
	give a contribution to $F$ in (30) 
	
	\begin{equation}
	(1/3)[-J^{a'a}_{i'i}(-\p',-\p)C^{cc'}(\r,\r')J^{bb'}_{ii'}(\q,\q')$$
	$$+J^{a'a}_{ij}(-\p',-\p)G^{cc'}(\r,\r')L^{bb'}_{j}(\q,\q')(-iR'_i)$$
	$$+(iR_j)L^{a'a}_{i}(-\p',-\p)G^{c'c}(-\r',-\r)J^{bb'}_{ji}(\q,\q')].
	\end{equation}

	\subsection{Contributions from Figs.15, 17}
	In this case, unlike the previous one, even when $\theta=0$, we cannot neglect ghost graphs, because only
	incomplete ghost graphs appear in Fig.17 (the 1 parts from Fig.5 without the $-P^2/\bar{P^2}$ parts);
	so we must take all the graphs. As explained above, they are taken with a coefficient
	$1/6$. 
	
	Thus the contributions have the form
	\begin{equation}
	\hbox{(1/6)[(ci)+(d}\gamma)'\hbox{+(ei)+(d}\alpha)\hbox{+(cii)+(eii)]}
	\end{equation}
	giving
	\begin{equation}
	\hbox{(1/3)[(ci)+(ei)+(cii)+(eii)]-(1/6)[(cii)+(eii)]}.
	\end{equation}
	For the first square bracket in (35), we use the identities in Figs.5,6 to get back to graphs like
	(c) and (e) in Fig.8, but with the $q$-line replaced by $\delta_{ij}$ and $1$ respectively. These are shown in Fig.19, and have a coefiicient $(1/3)$. This must be doubled because of graphs with the $p$- and $q$-lines
	intercanged. The result is (with the notation of (32))
	\begin{equation}
	(1/3)[2L_i^{a'a}(-\p',-\p)][\delta^{bb'}\delta(\q-\q')]
	 [iR_iG^{c'c}(-\r',-\r)-iR'_iG^{cc'}(\r,\r')].
\end{equation}

	There is also the contribution from the second square bracket in (35), with coefficient $(-1/6)$,
	but there is also a minus sign with (cii) and (eii) in Fig.15. This also must be doubled, giving
	\begin{equation}
	(1/3)[iP_iG^{aa'}(-\p',-\p)][\delta^{bb'}\delta(\q-\q')]
	 [iR_iG^{c'c}(-\r',-\r)-iR'_iG^{cc'}(\r,\r')].
\end{equation}

The expressions in (33), (36) and (37), inserted for $F$ in (30), constitute the generalization of the Christ-Lee terms when the external field is not transverse. 
\begin{figure}[h]
\centering
\includegraphics[width=0.8\textwidth]{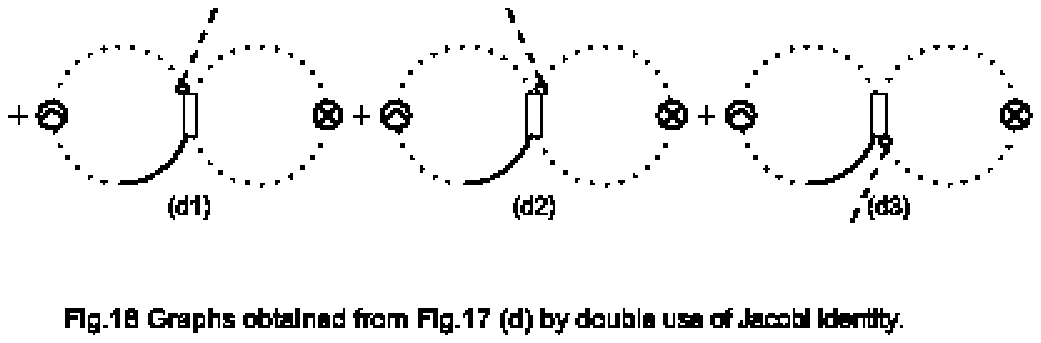}
\end{figure}
\begin{figure}[h]
\centering
\includegraphics[width=0.8\textwidth]{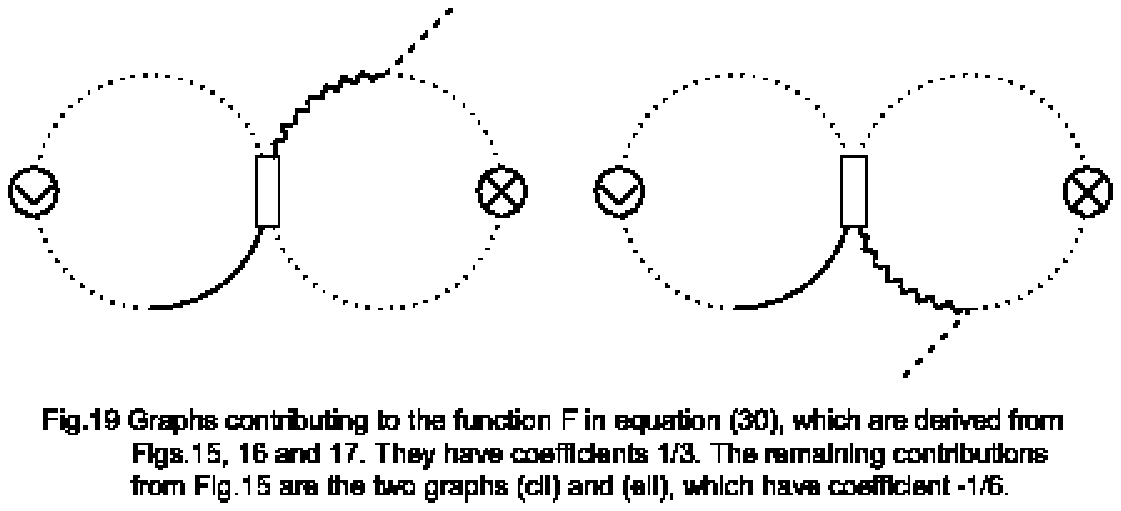}
\end{figure}
\begin{figure}[h]
\centering
\includegraphics[width=0.8\textwidth]{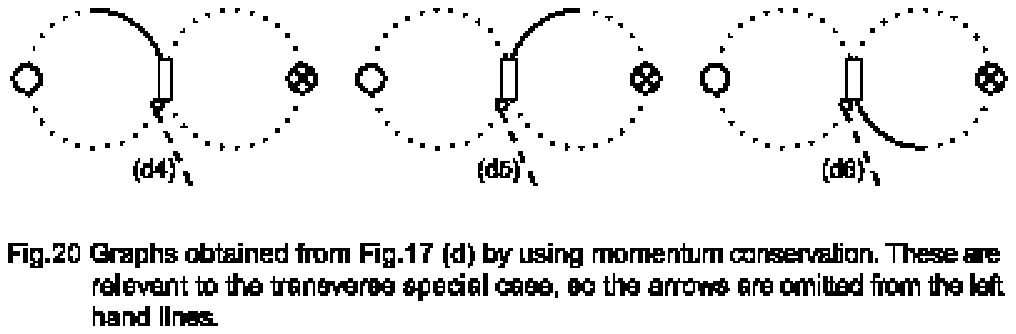}
\end{figure}
\section{The transverse special case}

In order to make contact with \cite{christlee}, \cite{doust}, we show how our result can be re-written
if the external gluon field is transverse. In this case, many simplifications occur. The covariant derivative $D$ obeys
$\partial.D=D.\partial$ and  so
\begin{equation}
G^{aa'}(\p,\p')=G^{a'a}(-\p',-\p),
\end{equation}
that is the arrows in Figs.3 and 6 and subsequent figures are not necessary.

We examine  the contributions to $F$ in equation (30) in this special case.
We note that in Fig.17(d), instead of using the Jacobi identity as we did to get Fig.18, we might alternatively use momentum conservation, which yields Fig.20. We will take the average of these two; that is Fig.18 with coefficient $1/6$ together with Fig.20 with coefficient $1/6$. Each of these contributions has to be doubled
because of the graphs with $p,q$ interchanged.

Then we take
\begin{equation}
\hbox{(1/3)[Fig.18 + Fig.15\{(ci)+(ei)\}]+(1/3)[Fig.20 + Fig.15\{(ci)+(ei)+(cii)+(eii)\}]}$$
$$=
\hbox{(1/3)Fig.19 \\\ +\\\ (1/3)Fig.21},
\end{equation}
where we have used the first and second identities in Fig.6. (We neglect terms like (28) which are zero in dimensional regularization.)

In \cite{doust}, equation (4.5.13), it is shown that the contributions from Fig.7(a),(b),(c) and Fig.8(a),(c) together with (39) are equal to  $V_1+ V_2$
of \cite{christlee}. Thus we have confirmed that,  in the transverse special case, our method agrees with the
known results.

\begin{figure}[h]
\centering
\includegraphics[width=0.8\textwidth]{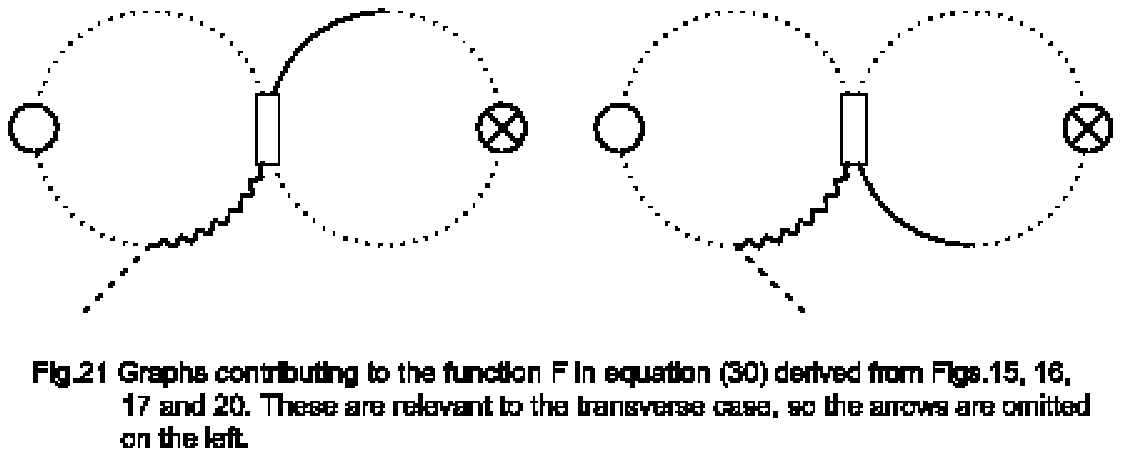}
\end{figure}
\section{summary and discussion}
We have proved that, in spite of energy divergences in individual 2-loop graphs, the effective
action in perturbative Coulomb gauge QCD is well defined. The sum of the relevant terms (in expressions (33), (36) and (37)), which is energy-independent, is
calculated, being a generalization of the function $V_1+V_2$ derived in a different way in
\cite{christlee}. These terms in the effective action would be a necessary contribution to the 
BRST identities. As a check, we have shown that in the special case of transverse external fields, our result
reduces to the known one.

What are the implications of the Christ-Lee function $V_1+V_2$ and its generalization in this paper?
There are two quite different approaches, the Christ-Lee approach \cite{christlee} and the Doust
approach \cite{doust}. The Former carefully studies the operator ordering in the Coulomb Hamitonian, and derives the Hamiltonian 
\begin{equation}
H_W+V_1+V_2
\end{equation}
where the suffix $W$ denotes Weyl ordering, which is necessary to avoid double counting. This is because of the zero vacuum expection value of
\begin{equation}
\{\hat{A}_i^a(\p)\hat{E}_j^b(\p)\}_W\equiv (1/2)\{\hat{A}_i^a(\p)\hat{E}_j^b(\p)+\hat{E}_j^b(\p)\hat{A}_i^a(\p)\},
\end{equation}
which implies that Feynman graphs like Fig.1, with pairs of operators contracted in the Coulomb potential operator, are omitted.

The Doust approach, followed in this paper, effectively defines the Coulomb gauge as the limit of the flow gauge given by equation (4), and there is no operator ordering problem in the flow gauge. 

Presumably the Christ-Lee Hamiltonian (40) might be used in Hamiltonian lattice QCD, but the Doust approach
would not be appropriate. On the other hand, the effective action for non-transverse fields
cannot be derived in the Christ-Lee approach; and this general effective action is required to
implement BRST identities.

 \section*{Appendix - Definition of effective action}
The effective action is normally defined by first introducing a source current $J$:
\begin{equation}
\int d^dx J^a_i(x)\hat{A}^a_i(x)
\end{equation}
(for simplicity, we omit sources for other fields). Then the external classical field $A^a_i(x)$ is introduced by a Legendre  transformation from $J$ to $A$. If the transversality constraint is imposed by a delta function:
\begin{equation}
\delta\{\hat{A_i^a}\}
\end{equation}
there appears to be a problem, because then  only transverse $J^a_i$ contribute in (42).

To avoid this, we may impose transversality by using a gauge-fixing term like
\begin{equation}
\frac{1}{2\theta^2}\int d^dx \{\partial_i \hat{A}_i^a(x)\}^2
\end{equation}
and taking the limit $\theta \rightarrow 0$ only after the effective action has been defined.
Actually, for the purpose of the present paper, we should use the gauge fixing term in equation (4) in place of (44).




\end{document}